# Human-AI Collaboration for Wearable Technology Component Standardization

Andrew M. Lydner
Georgia Institute of Technology alydner3@gatech.edu

***Abstract***

Due to the multidisciplinary nature of wearable technology, the industry faces potential limitations in innovation. The wearable technology industry is still in its infancy and increased applicable use faces stagnation despite the plethora of technologies that have been largely wrist worn. This could be a result of the lack of multidisciplinary expert knowledge disseminating through the industry. Unlike other technologies which have standardizations and processes for how they are developed, wearable technologies exist in a realm of perpetual change as given the various materials and subcomponents that continue to be developed. It is essential that expert opinions form a collaborative foundation, and even more so that intelligent systems foster that collaboration. The caveat though, is likeliness of these artificial intelligence (AI) collaboration tools to be utilized by industry experts. Mental model development for AI tool usage could be applied to wearable technology innovation in this regard, thus the goal of this paper and focus of research.

## 1. INTRODUCTION

Wearable technology innovation is highly dependent on collaborative efforts and must incorporate the use of intelligent systems for knowledge synthesis from diverse viewpoints to prevent sporadic-researched and unstandardized technology. The wearable technology industry has developed within the realms of several generations of computing, however since the inception of its modern-day flagship items from wearable wristbands such as smartwatches and activity trackers, there has been little to no consensus on the wearability of everyday garments and other applicable wearable items that are used in industries such as healthcare largely due to the multidisciplinary nature of wearable technology's innovation. Research and development for wearable technology takes a multifaceted approach from a wide range of industry experts. Convergence on standards of materials and subcomponents that are practical, electronically sound, and human factored are of the essence. " Like humans, AI makes mistakes, and humans can catch and compensate for AI errors" (Ren et al., 2023). This paper delves into aspects of educational technology research that can be used to incite innovation in wearable technology through collaboration between AI and experts within the multidisciplinary field. Through pedagogical and experimental research, efficacy of specific methods could result in standardization for various existing and future wearable technologies. This paper delves into research that tests an initial hypothesis that educational courses while leveraging pedagogical research from social learning, the Hawthorne effect, and other learning principles develops students' mental model for AI tool use. The goal is to determine if the result of mental model development from such courses result in actual use of multidisciplinary collaboration tools specially for wearable technology. "Human+AI predictions were more accurate than AI predictions across all conditions." (De-Arteaga et al., 2023).

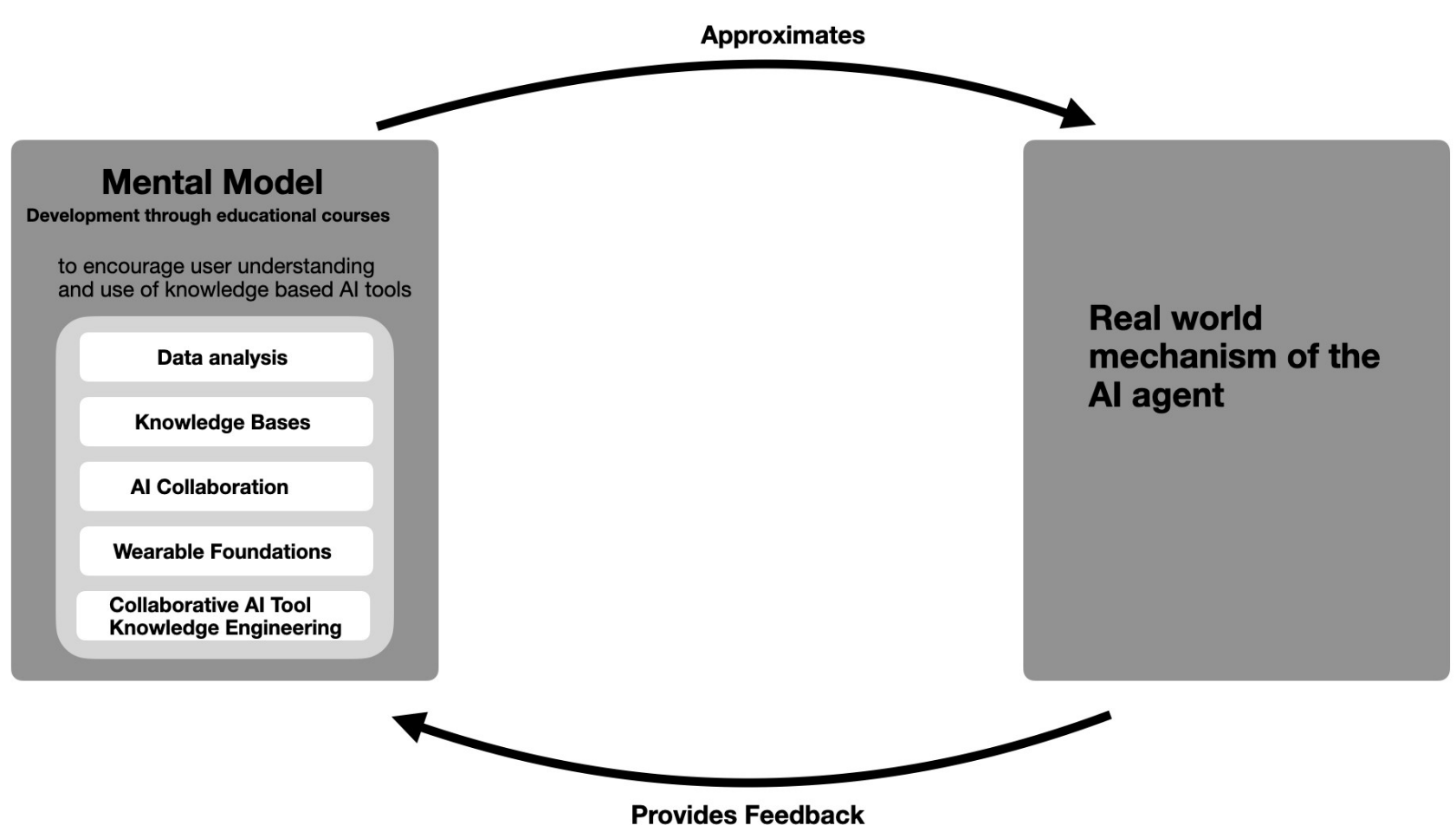

***Figure 1***—Make sure your flowcharts are more useful than this one. Source: XKCD.

## 2. METHODOLOGY

To conduct this research, an educational course that teaches mental model development and an AI collaboration tool which comprises an agent that reasons over expert input knowledge was developed simultaneously. A controlled study was then carried out to track user activity with the AI collaboration tool following course completion. Both groups in the experiment received images as a representation of a mini course and were asked to traverse through them on a survey platform that confirmed their understanding for each module in the course. The various modules of the course included data analysis, knowledge bases, wearable foundations, AI collaboration, and multidisciplinary collaborative AI tool knowledge engineering. Following the course module traversals, survey respondents were given login credentials to access the developed tool. In the experiment, the experimental group received course representation with image notification of exclusive access to the developed AI collaboration tool upon completion. The control group received images that had no explicit notification of AI tool access. Both groups in the experiment were tested for their login rate and the rate that they entered data in the tool to collaborate with the AI agent.

## 3. RESULTS

The results of the experiment confirmed the hypothesis that educational courses could be an integral component in the AI tool uptake pipeline, specifically driven by educational courses that leverage the Hawthorne effect to incite mental model development. As shown in Figure 2, there was a small percentage of users from each group that input into the tool, however 64.7% of all the learners that completed the course are from the experimental group. This suggests that the Hawthorne effect motivates students to complete courses related to mental model development.

| | **Control group** | **Experimental group** |
|---|---|---|
| **Survey respondents** | 12 | 22 |
| **Users who entered data in the tool** | 1 | 1 |
| **Rate of course completion as a percentage of all learners in study** | 35.2% | 64.7% |
| **Rate of course completion leading to tool usage** | 8.3% | 4.5% |

***Figure 2***—The experimental group had higher course completion rates

## 4. RELEVANCE TO HUMAN FACTORS THEORY

Wearable technology's prevalence coincides with the societal expectations and reliance on the internet-of-things. Standardization for wearable technologies is necessary due to the high value and innovation that results from a wearable technology's successful development and successful use of such technologies by human beings. The standardization of components for wearable technology results in adequate knowledge resources distributed for very specific items which informs the human-centered design.

In a study exploring longitudinal physiological monitoring among older adults with mild cognitive impairment, researchers deployed a smart ring for health monitoring proposes. The researchers concluded that both physical and psychological conform is important to ensure device user satisfaction. Many of the participants complains about the discomfort of the rings and the inability to use them due to physical constraints. The researchers notes that "Arthritis and swelling were a primary concern for participants when using this device. Despite using the size kit before starting the study, participants often experienced swelling in their fingers, impacting their comfort when taking it on and off or wearing it throughout the day" (Gleaton et al., 2024). The experiences of the participants demonstrate the importance and relevance of human factors to wearable technology component standardization. It's evident that had there been a collaborative effort in wearable technology component standardization for the smart ring used in the study, specifically with emphasis on the right materials to use given human physical constraints, participant satisfaction would have become more apparent. Thus, higher quality research in the interdisciplinary field of wearable technology could disseminate and further develop the industry.

## 5. WEARABLE TECHNOLOGY AND STANDARDIZATIONS

The wearable technology industry comprises a vast area of research from computational materials and textiles to products that are applied in healthcare, and industry. Standardization for the real-world applicability of such items is crucial to spurring innovation as a lack of multidisciplinary knowledge could result in knowledge gaps in the specific industry use cases and requirements for each item.

## 6. COLLABORATIVE ARTIFICIAL INTELLIGENCE FOR MULTIDISCIPLINARY INDUSTRIES

Artificial intelligence algorithms, specifically agentic AI as knowledge based artificial intelligence and cognitive systems algorithms are prime candidates for the wearable technology industry due to the nature of knowledge bases and expert opinions. Knowledge-based AI is based on inferences and reasoning over a set of criteria, which in this case is entered by various experts. As seen in in figure 3, cognitive computing paradigms such a natural language processing plays an integral part in synthesizing the knowledge before storing and reasoning over this into a knowledge base. Moreover, each user that inputs knowledge into the knowledge base, based on their credential could be given a weight which is factored into how their inputs is a part of the output of the collaborative-driven result.

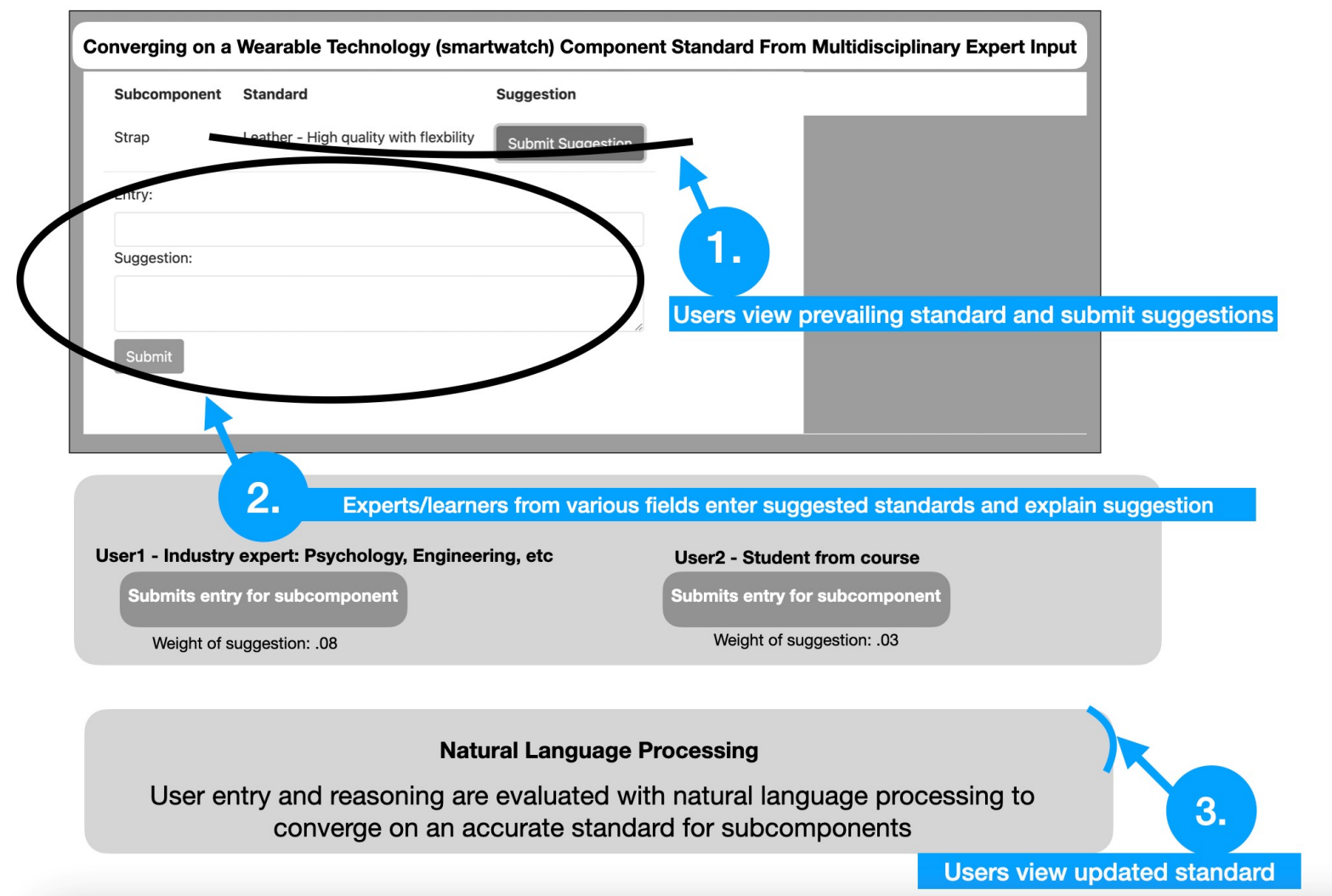

***Figure 3***—User inputs are represented and reasoned over with natural language pro-

## 7. THE NECESSITY OF MENTAL MODELS FOR ARTIFICIAL INTELLIGENCE COLLABORATIVE TOOLS

Artificial Intelligence tools have been traditionally seen as having algorithms that rely on a black box so to speak, with users not exactly sure of how the result came to be. With properly organized knowledge-based AI tools, the result is explicitly stated with adequate reasoning for each result. Mental models for knowledge based artificial intelligence tools surrounds the users understanding how and why the result came about. Mental model development for artificial intelligence results in usable tools that can ensure collaboration among experts from various fields, a necessity when developing innovation and standards for an industry such as wearable technology. Moreover, with an educational-focused methods of encouraging the actual use of industry-based AI tools, such as research-based pedagogy, users of AI tools could find such systems easier to engage with and much more approachable than they would have otherwise.

## 8. EDUCATIONAL COURSES AS A MODE OF SPURRING AI COLLABORATION TOOL USE

With the rise in Artificial Intelligence tools such from generative AI that leverage machine learning to assist experts with daily tasks such as knowledge management, often they aren't aware of how results pertaining to algorithmic outputs are derived. Educational courses that focus specifically of developing mental models for industries based on how those tools will be used could be pivotal in significant use by industry experts, which in the case of wearable technology, could result in increased innovation, dissemination of expert knowledge, and standardization for components of the various items under development.

## 9. EXPERT KNOWLEDGE CONVERGING ON STANDARDS

Expert knowledge is essential to the continuous integrity and reliability of knowledge-based systems which leverage cognitive computing to set standards in multidisciplinary industries. In the developed AI collaboration tool, inputs by experts are organized into a knowledge base and compared against existing entries by a process of candidate elimination or version spaces. Given the user type and industry, specific weights leverage exactly how the algorithms interpret and organizes the inputs. This new knowledge from experts reinforces criteria that exists or doesn't exist in the pool of knowledge. Regarding wearable technology, while students who develop mental models are certainly stakeholders capable of using the tool, as seen in the experiment, seasoned experts with credentials from various fields hold more expert knowledge and capability to disseminate knowledge in such an ever-changing field. "Results suggest that complementarity is easier to achieve when loss rates are highly variable: when the unaided human (or algorithm) has very low loss on some inputs and very high loss on inputs. Disparate levels of loss raise issues of fairness" (Donahue et al., 2022).

### 9.1. Psychologists and Human Factor Engineers

Psychologists, human factor engineers, and ergonomics experts serve as some of the main users of an AI tool that synthesizes expert knowledge to generate wearable technology standards. Attributes such as a wearable materials' comfort and propensity of certain materials to result is fatigue or frustration are all factors that are studied by psychologists and ergonomics experts. Thus, inputs and insights form such experts are invaluable to AI Collaboration tools for wearable technology.

### 9.2. Human-Computer Interaction Engineers

Human-Computer Interaction engineers have unique understanding of human interactions with technological devices. Expert knowledge that encompasses an understanding of user goals and user needs is the hallmark of considerations for weightage of inputs in AI collaboration tools for wearable technology. With a plethora of computational materials and other components experimented with for user applicability, it’s imperative that a fundamental understanding of the advantages and disadvantages of the various materials affect the user and which of those subcomponents dissuade or encourage proper use of wearables in the way they were intended.

### 9.3. Industrial Engineers

Industry requires rigorous study and understanding of developing and manufacturing wearable technology devices. Industrial Engineers are prime in their ability to discern which materials and components fit within a set of constraints given the intentions of the products’ designers.

### 9.4. User Interface Designers and Researchers

Every interaction with a technological device is an opportunity for user interface designers to iterate on the design, and carryout research to test and evaluate the efficacy of such interfaces. User Interface designers and researchers’ inputs into AI collaboration tools for wearable technology serve as fundamental knowledge in the knowledge base for multidisciplinary collaboration, large due to the extensive embassies of design and research that’s user centered.

### 9.5. Fashion and Garment Designers

Wearable technology development cannot exist in a space without the input of experts in the field of fashion and garment design due to the prototyping expertise as it relates to the fashion industry. Moreover, fashion and garment designers’ inputs’ can leverage their expertise and knack for accessing market needs as certain materials are suitable for certain industries. For example, swimwear is often made of using elastane, polyester, and nylon. Only a fashion designer would be aware of the appropriate materials needed for this. Fashion and Garment designers have experience in manipulating garments to suit the correct user.

### 9.6. Textiles and Fiber Science Engineers

Textiles and fiber science engineering has grown to involve the study of computational materials. Most existing wearable technology devices leverage materials that have already been standardizes for their non-technological peers such as cotton for clothing, rubber or leather for smartwatch wristbands, and Steel for a smartwatch’s hardware. Research has shown that computation can be embedded into the fiber of certain materials. This gives rise to a new paradigm of wearable technology as the industry shifts away from standards solely by experts in one specific field. Textile and fiber science, regarding wearable technology often intertwines with electrical engineering to converge on and understand how computational materials be washable, wearable, retain electrical integrity and safety, and even manufactured at scale.

### 9.7. Healthcare Practitioners

Most of the existing uses of wearable technology that exist within the realm of healthcare are dependent on activity tracking and health monitoring. Healthcare practitioners have deep insight into the existing products that are useful in clinical settings. Thus, the importance of their expert opinions in determining which wearable technology subcomponents suffice for specific items.

### 9.8. Electrical Engineers

Electrical engineers traditionally dealt with electrical components regarding wiring, safety, and integrity. Wearable technology development requires knowledge from electrical engineers, specifically due to their understanding of how certain wires and materials function within a given context. Moreover, certain materials for items are heat resistant and may interact with electrical wiring. Thus, the necessity of electrical engineers to not only validate and reinforce inputs from experts, but also to disseminate their knowledge within the context of what has been entered from previous entries considering the difference in weightage.

## 10. THE USE OF KNOWLEDGE-BASED AI COGNITIVE COMPUTING AS AN INNOVATION TOOL TO DEVELOP WEARABLE TECHNOLOGY IN VARIOUS INDUSTRIES

The developed wearable technology collaboration tool's AI agent reasons over inputs from experts in multidisciplinary fields. At the center of the reasoning of the inputs is natural language processing technology and its associated deductive reasoning properties. As seen in figure 4, the pipeline from input to output is largely dependent on natural language processing and knowledge representation and reasoning version spaces. The system fosters innovation within wearable technology by leveraging expert inputs to set a standard for various items given weightage based on expert credentials. The joining of knowledge and reasoning from experts and Artificial Intelligence results in a combining of previous experience and expert .

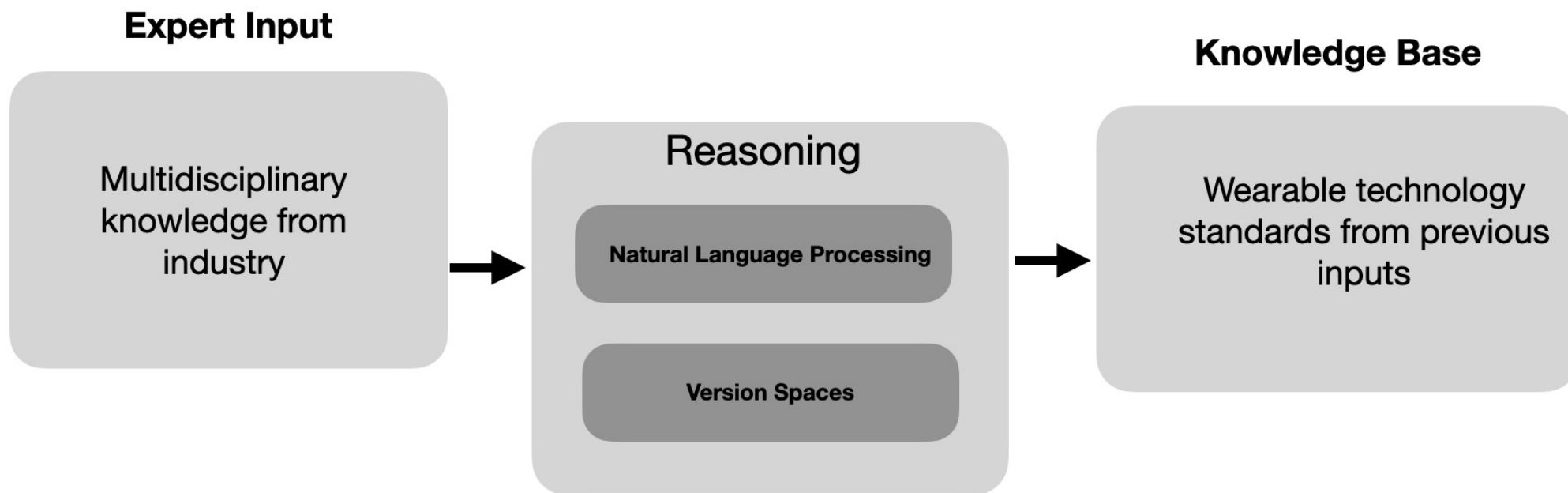

***Figure 4***— Knowledge is reasoned over with natural language processing and version spaces

In a study exploring the reliability of modern AI paradigms such as ChatGPT, researchers concluded that a reasoning aspect leveraging knowledge bases and deductive reasoning is essential to ensure more accurate results compared to the common efficient AI which generally takes any data as ground truth and could be seen as a black box. In proposing an knowledge-based AI architecture which could be applied to and improve AI based tools such as ChatGPT, researchers noted that "core of this architecture lies in integrating knowledge reasoning-based hybrid computing, numerical computing, symbolic computing, data mining, program verification, mathematical logic, and other technologies with data-driven techniques such as annotation, understanding, recognition, and generation of text, audio, and image data. This approach not only ensures the efficiency of AI systems but also enhances the reliability of AI applications" (Liu and Qin, 2024). The study's success in deploying knowledge based-ai architecture as an enhance reasoning-based agent demonstrates the advantages of human-ai collaboration. Reasoning and synthesis of knowledge from experts cab further refine and improve AI tool accuracy, which int he case of wearable technology components, guarantees more precise materials which could enhance quality and user satisfaction.

## 11. CASE BASED REASONING IN KNOWLEDGE BASED AI FOR THE WEARABLE TECHNOLOGY INDUSTRY

Cased-based reasoning is applying solutions from old problems to new problems. A knowledge-based artificial intelligence agent can use case-based reasoning to apply expert-agreed standards from one industry to another. The way that this is useful in the wearable technology industry or wearable-technology adjacent industry, is so that standards can be set in areas that are largely independent from wearable technology. For example, in the case of manufacturing, certain components that are used in smartwatches or smart apparel may be applied to a smart factory such as the type of metal on the levers or even the kind of materials, computational and otherwise that may be suitable for that use case.

## 12. KNOWLEDGE REPRESENTATION AND REASONING FOR STANDARDIZATION

Knowledge representation in knowledge bases for standardization is useful for expert collaboration, especially for setting standards and fostering the decision-making process. Standardization in wearable technology requires the organization of unstructured knowledge. To effectively converge on a standard due to a series of user inputs, reasoning must be systematized.

## 13. LIMITATIONS

The developed application poses significant opportunities for collaboration among experts in multidisciplinary fields, however, there are some limitations that arise. An imbalance of experts from various specific fields, the overabundance of one type of expert, and the lack the AI reasoning capabilities due to insufficient industry knowledge within the knowledge base. These limitations can be overcome through careful weightage of expert opinions and human-centered algorithms.

## 14. CONCLUSION

Wearable technology is a field whose development is highly dependent on expertise from various industries from ergonomics and human factors to fashion design and electrical engineering. To effectively disseminate knowledge pertaining to wearable technology subcomponents, users of collaborative artificial intelligence tools must have mental models of what to expect from such tools and how the reasoning and decision-making process was delivered. Educational courses focusing on pedagogy from education research related to mechanisms that encourage the completion of courses related to mental model development could result in marked uptake of AI tools for wearable technology. Innovative development and embedding of internet of things devices and computational materials require a heed of collaborative expert decision making from various industries, only then standardization can be achieved and innovation in wearable technology items can be as prevalent as traditionally worn items.